\documentclass[aps,prb,twocolumn,superscriptaddress,showpacs]{revtex4}

\newcommand \bmat {\begin{displaymath}}
\newcommand \emat {\end{displaymath}}

\newcommand{\tw}{t_{\rm w}}
\newcommand{\twone}{t_{\rm w1}}
\newcommand{\twtwo}{t_{\rm w2}}

\newcommand{\tmerge}{t_{\rm trans}}

\newcommand{\Ts}{T_{\rm s}}
\newcommand{\ts}{t_{\rm s}}

\newcommand{\teff}{t_{\rm eff}}
\newcommand{\teffcum}{t_{\rm eff}^{\rm cum}}

\newcommand{\Tg}{T_{\rm g}}

\newcommand{\Ti}{T_{\rm i}}
\newcommand{\Tm}{T_{\rm m}}

\newcommand{\kb}{k_{\rm B}}

\newcommand \be {\begin{equation}}
\newcommand \ee {\end{equation}}

\newcommand \bi {\begin{itemize}}
\newcommand \ei {\end{itemize}}

\def\nle{\ \raise.3ex\hbox{$<$}\kern-0.8em\lower.7ex\hbox{$\sim$}\ }
\def\nge{\ \raise.3ex\hbox{$>$}\kern-0.8em\lower.7ex\hbox{$\sim$}\ }

\usepackage{graphicx}
\usepackage{epsf}
\usepackage{amssymb}
\usepackage{array}

\begin{document}

\title{Absence of rejuvenation in a Superspin Glass}

\author{P. E. J{\"o}nsson}
\affiliation{ISSP, University of Tokyo, Kashiwa-no-ha 5-1-5, Kashiwa, Chiba 277-8581, Japan}

\author{
H. Yoshino
}
\affiliation{Department of Earth and Space Science, Faculty of Science,
Osaka University, Toyonaka, 560-0043 Osaka, Japan}

\author{H. Mamiya}
\affiliation
{National Institute for Materials Science, Sengen 1-2-1, Tsukuba, Ibaraki 305-0047, Japan
}

\author{H. Takayama}
\affiliation{ISSP, University of Tokyo, Kashiwa-no-ha 5-1-5, Kashiwa, Chiba 277-8581, Japan}

\date{\today}

\begin{abstract}

Effects of temperature changes on the nonequilibrium spin-glass dynamics of a strongly interacting ferromagnetic nanoparticle system (superspin glass) are studied.
In contrary to atomic spin glasses, strong cooling rate effects are observed, and no evidence for temperature-chaos is found.
The flip time of a magnetic moment is much longer than that of an atomic spin and hence much shorter time scales are probed within the experimental time window for a superspin glass than for an atomic spin glass. Within a real space picture the cumulative aging observed for the superspin glass can be explained considering that all investigated length scales are shorter than the temperature-chaos overlap length.
The transient relaxation, observed in experiments after temperature changes, can be understood as the adjustment of thermally active droplets, which is  mutatis mutandis the Kovacs effect observed in most glassy systems.

\end{abstract}

\maketitle

\section{Introduction}

Spin glasses have been an active field of research since the 70th's, yet many questions about the nonequilibrium dynamics in the spin-glass phase are still unsolved. 
Magnetic aging in spin glasses was first observed in 1983 by Lundgren et al.\cite{lunetal83} 
Physical aging was already familiar in the field of structural glasses \cite{struik78} and later many different types of materials have been shown to exhibit aging in various physical quantities,  e.g. strongly interacting nanoparticles \cite{jonetal95}, granular superconductors \cite{papetal99}, oriental glasses \cite{albetal98},  and gels.\cite{cipetal2000}
More intriguing is the sensitivity to temperature changes that has been observed experimentally for spin glasses; after a sufficiently large temperature change the system appears unaffected by previous aging, i.e, it is {\em rejuvenated}.\cite{gralunnor90,ghoststory}

In contrast to experiments, rejuvenation effects can hardly be detected in numerical MC simulations on the Edwards-Anderson Ising model.\cite{takhuk2002}
The different sensitivity to $T$-perturbations for numerical simulations and experiments can be attributed to the different time scales (in unit of the microscopic flip time $\tau_m$) investigated in the two cases ($\tau/\tau_m \in [1,10^6]$ for numerical simulations while $\tau/\tau_m \in [10^{12},10^{18}]$ for experiments on atomic spin glasses).
Superspin glasses (strongly interacting nanoparticle systems) are therefore interesting to investigate since the experimental time window corresponds to shorter time scales than for atomic spin glasses, due to the longer microscopic flip time of a superspin (magnetic moment) compared to that of an atomic spin.
In several works strongly interacting nanoparticle systems, such as frozen ferrofluids, have indeed been shown to exhibit spin-glass dynamics.\cite{jonetal95, mamnakfur99,jon2004} In those systems the dominant interparticle interaction is the dipolar interaction. Disorder and frustration appears due to the randomness in the placement of the particles and in the directions of the anisotropy axes.

In this article we will investigate the effects of temperature changes of the nonequilibrium dynamics of a sample consisting of a solidified ferrofluid, in which nanoparticles of Fe$_3$N are randomly distributed.
The results will be compared with atomic spin glasses and other glassy systems.
Spin glass properties of this sample has earlier been investigated in Refs.~\onlinecite{mamnakfur98,mamnak99,mamnakfur99,jonmamtak2004}.

\section{Background}
\label{sec-back}
We will briefly discuss the effects of temperature changes on the nonequilibrium dynamics of a spin glass based on a real-space picture.\cite{fishus88} 
For simplicity, the discussion is made for an atomic Ising spin glass with nearest neighbor interactions, but it is also valid for systems with long-range interactions such as the dipolar interaction. To a first approximation superspins with uniaxial anisotropy can be described as Ising spins. 
The nonequilibrium dynamics, at temperature $T$ below the spin-glass
transition temperature $\Tg$, is governed by the slow growth of
thermally activated droplets of linear size $L_T(t)$. 
Due to randomness and frustration a temperature change will affect the nonequilibrium dynamics in the following two respects:
the population of the thermally active droplets  on short length scales
is strongly temperature dependent, and
temperature-chaos causes strong
rejuvenation effects on the length scales larger than  the so called 
overlap length $L_{\Delta T}$. The latter is defined so that
equilibrium spin configurations at temperatures $T$ and $T+\Delta T$ are
completely uncorrelated on length scales larger than $L_{\Delta T}$.
The dominating effect depends on the investigated length scales with
respect to $L_{\Delta T}$ corresponding to the temperature change.

Here we want to emphasize the difference between rejuvenation
effects originating from the temperature-chaos nature of spin-glass
states and other effects that are not related to it.
It will be argued that effects not related to $T$-chaos are transient effects since they vanish within a finite time, but may
give rise to ``apparent rejuvenation'' within a finite time window.
Such apparent rejuvenation, involving length scales less
than $L_{\Delta T}$, was called  ``fixed energy-landscape
rejuvenation'' in Refs.~\onlinecite{bouetal2001,berbou2002}. 
Let us consider a
negative temperature-shift $T+\Delta T \to T$ with $\Delta T >0$. By the
1st aging for a time $\tw$ after a temperature quench down to $T+\Delta
T$, the system is equilibrated up to a certain length scale $L_{T+\Delta
T}(\tw)$. 
After the temperature-shift ($T$-shift) the aging continues at the new temperature $T$  as if the system had been aged at this temperature for a certain effective time $\teff$.
If this $\teff$ is equal to $\teffcum$, defined by
\be
L_{T}(\teffcum)=L_{T+\Delta T}(\tw),
\label{eq-acc}
\ee
the process is called {\em cumulative aging}; spin-glass domains of mean size $L_{T+\Delta T}(\tw)$ grown up in the 1st aging at
$T+\Delta T$ remain at the new temperature $T$.
For large $\Delta T$,
however, it may occur  that $L_{T}(t_{\rm eff})$ is smaller than
$L_{T+\Delta T}(\tw)$. This implies the existence of another length scale
which is less than $L_{T+\Delta T}(\tw)$, and above which the locally
equilibrated configuration with respect to $T+\Delta T$ is irrelevant to
the aging dynamics at $T$. That length scale is the overlap length
$L_{\Delta T}$. Therefore a crucial check of the rejuvenation due to
temperature-chaos is to observe whether Eq.~(\ref{eq-acc}) is violated or
not.

In Sec.~\ref{sec-T-change} we show
that temperature-chaos is not relevant in the present sample for the
temperature changes we can investigate within the experimental time
window. Therefore, in the following, we will focus on transient effects not related to $T$-chaos.
To this end, we focus on the temperature
dependence of the population of thermally active droplets in cumulative
aging, assuming that the investigated length scales are much shorter than
$L_{\Delta T}$.

In spin glasses the surface 
free-energy $F_{L} (>0) $ of a droplet of size $L$ strongly fluctuates
between different droplets at different places in the system.
The {\it typical} surface free-energy scales as $\Upsilon(T) (L/L_{0})^{\theta}$ 
with the stiffness exponent $\theta >0$ and $\Upsilon(T)$ being the stiffness constant which is constant at low temperature ($T \ll  T_{\rm g}$) but vanishes as $T \to T^{-}_{\rm g}$.
There are
marginal droplets with $F_{L}$ smaller than the thermal energy $\kb T$
with non-zero probability, and 
thus the probability
$p_{L}(T)$ that a given Ising spin belongs to a thermally activated 
droplet of linear size $L$ at temperature $T$ is given as,\cite{komyostak2000,komyostak2000A}
\be
p_{L}^{\rm SG}(T) 
\sim  \tilde{\rho}(0)\frac{\kb T}{\Upsilon(T) (L/L_{0})^{\theta}}\,.
\label{eq-prob}
\ee
The first important point is that the probability exhibits a {\it linear}
temperature dependence.  Hence the amount of droplets which must be 
deactivated (turned off) after negative $T$-shift $T+\Delta T \to T$
or activated (turned on) after positive $T$-shift 
is proportional to the magnitude of the temperature difference $\Delta T$.
The second important point is that the probability only decays
algebraically with increasing length $L/L_{0}$. One should also recall
that the stiffness exponent is known to be very small $\theta \approx 0.2$ 
in 3-dimensional Ising spin-glass.\cite{bray84,hartmann99,komyostak99} 
We can compare to the case of a multi-domain ferromagnet for which the probability that a given Ising spin belongs to a thermally active
droplets of linear size $L$ at temperature $T$ is given as,
\be
\qquad p_{L}^{\rm FM}(T) 
\sim \exp \left( - \frac{\Upsilon(T) (L/L_{0})^{d-1}}{\kb T} \right),
\ee
where $d$ is the spacial dimension.
In this case, the existence of thermally active droplets becomes exponentially small at macroscopic scales $L/L_{0} \gg 1$. Any change of the population of
such rare droplets will hardly be seen in practice.

Now let us consider the negative $T$-shift $T+\Delta T \to T$ above introduced, for which $\teff \simeq \teffcum$ holds. 
The characteristics of the spin configuration just after the temperature
change is essentially the same as that would be obtained after an aging
performed directly at $T$ for a period $t_{\rm eff}$. However it is
important to notice that certain amount of thermally active droplets are
{\it excessively active} with respect to the equilibrium state at the new temperature $T$. 
The deactivation of a droplet at the new temperature $T$ has a probability 
\be
p_{L}^{\rm SG}(T+\Delta T)-p_{L}^{\rm SG}(T) \sim 
 \tilde{\rho}(0)\frac{\kb \Delta T}{\Upsilon(T) (L/L_{0})^{\theta}}.
\ee
This deactivation is a dynamical process which takes a certain macroscopic time.  
At time $t$ after the $T$-shift, 
droplets are correctly thermalized with respect to the new temperature
$T$ only up to $L_{T}(t)$ and excessively active droplets are still present on 
larger length scales. Such large scale droplets will act as 
a frozen-in domain walls on the shorter time scales of smaller droplets.

Here let us recall the arguments by Fisher-Huse on the relaxation of the
ac susceptibility.  The ac susceptibility of a certain angular frequency $\omega$ 
probes {\it effective stiffness} of a droplet of size $L_{\omega}=L_{T}(1/\omega)$.
In the presence of a large-scale frozen-in domain wall of size $L$,
a small droplet of size $L_{\omega}$ will happen to stay in the vicinity of
the domain wall with a certain probability.  The free-energy cost needed to
activate a small droplet which touches the domain wall will be smaller
than those in the bulk. Thus the ac susceptibility will have a certain 
excessive contributions $\delta \chi(L_{\omega},L)$ which vanishes only
in the limit  $L/L_{\omega} \to \infty$.

Now combining the above ingredients we can calculate the ac
susceptibility $\chi_{\omega}(t;T)$ at time $t$ after the negative $T$-shift as,

\begin{eqnarray}
\label{eq-trans}
\chi_{\omega}(t;T)&=&  \chi^{\rm ref}_{\omega}(t+t_{\rm eff};T) \\
&+&\int_{L_{T}(t)}^{L_{T}(\tmerge)}\frac{dL}{L} 
\tilde{\rho}(0)\frac{\kb \Delta T}{\Upsilon(T) (L/L_{0})^{\theta}} 
\delta \chi(L_{\omega},L). \nonumber
\end{eqnarray}
Here $\chi_{\omega}^{\rm ref}(\omega)(t;T)$ is the reference curve which can be obtained by a direct quench to $T$ and $\tmerge \approx \teff$ is the time required to turn off excessively active droplets.
The 2nd term in r.h.s is the transient part of the susceptibility due to frozen-in excessively active droplets at length scales larger than $L_{T}(t)$, which vanishes after the time  $\tmerge$.
If one  only measure the ac susceptibility after the $T$-shift on time scales shorter than $\teff$, the isothermal and the transient contribution to the susceptibility  can not be separated and the susceptibility {\em appears} rejuvenated according to the ``fixed energy-landscape rejuvenation'' picture.

In the case of positive temperature-cycling $T \to T+\Delta T \to T$
(which is discussed in Sec.~\ref{sec-T-change}) the transient susceptibility after the $T$-cycling will also be described by Eq.~(\ref{eq-trans}).

\section{Investigation of the sensitivity to temperature changes}
\label{sec-T-change}

In order to make a quantitative analysis of the effect of $T$-changes on the aging, it is desirable to make experiments with fast temperature changes. 
Instantaneous temperature changes can unfortunately never be performed experimentally. 
The maximum cooling/heating rate of the MPMS-XL magnetometer used for our experiments was 10~K/min while the time needed in order to stabilize the temperature was $\sim 100$~s.
In the following, a ``temperature quench'' will refer to a cooling using this maximum cooling rate. The maximum cooling/heating rate is used in all experiments unless otherwise specified.

The Fe$_3$N nanoparticle system exhibits spin-glass dynamics below $T_g \sim 60$ K.\cite{mamnak99}
In all experiments $T=120$~K is chosen as the reference temperature at which the sample exhibits no slow dynamics.

\subsection{Cumulative aging}

\begin{figure}[htb]
\includegraphics[width=0.45\textwidth]{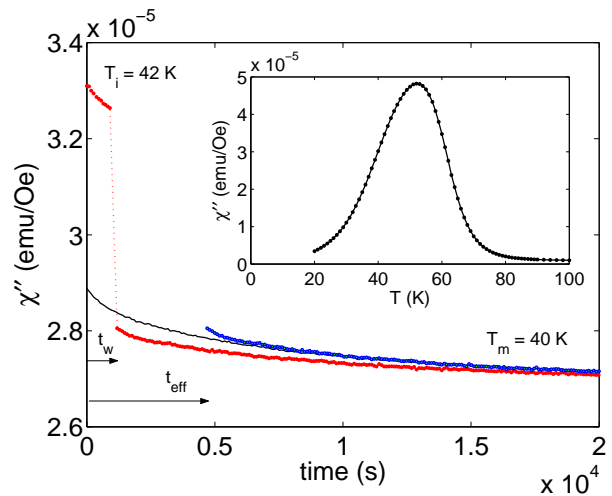}
\caption{(Color online) ac susceptibility vs time. The sample is aged at $\Ti =42$~K for a time $\tw$ before the temperature is shifted to $\Tm=40$~K. The solid line show the isothermal reference curve at 40~K. The effective age $\teff$
after the $T$-shift can be determined by shifting the ac-susceptibility at $\Tm$ by an amount  of time $\teff$ in order to make it fall on top of the reference curve at times larger than $\tmerge \approx \teff$.  
The inset show the ac susceptibility vs temperature. 
$\omega/2\pi = 510$~mHz and $h=1$~Oe.
}
\label{ac-shift}
\end{figure}

In a $T$-shift experiment, the sample is quenched from a high temperature to a low temperature $\Ti$ at which the sample is aged for a time $\tw$ before the temperature is changed to $\Tm$ and the magnetization (ZFC or ac) is recorded at the new temperature.
In order to quantify the effect of the aging at $\Ti$ at the new
temperature $\Tm$ an {\it effective} waiting time $\teff$ can be determined from the measurements. 
Since the ac-susceptibility relaxes monotonously under isothermal aging, the level of the ac susceptibility reflects directly the age of the system. 
The effective age $\teff$  of the system at $\Tm$ can be determined by shifting the ac curve so that it falls onto the isothermal aging curve at large time scales (see Fig.~\ref{ac-shift}).\cite{note1FeN}

\begin{figure}[htb]
\includegraphics[width=0.5\textwidth]{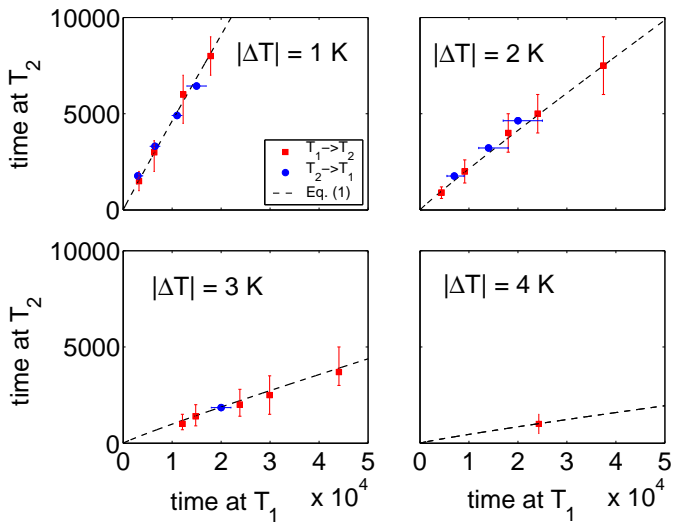}
\caption{(Color online)
$t_1$ ($\teff$ or $\tw$) at $T_1 = 40$~K~$- \Delta T$ vs $t_2$ ($\teff$ or $\tw$) at $T_2 = 40$~K. If the aging is cumulative, the two sets of data from positive $T$-shifts ($T_1 \to T_2$) and from negative $T$-shifts ($T_2 \to T_1$) will fall on the same curve corresponding to the domain growth law. The dashed lines indicate a separation of times with temperature according to Eq.~(\ref{t2}) with $\tau_0 = 4\cdot 10^{-10}$~s.
}
\label{teff}
\end{figure}

It is possible to check whether the aging is cumulative or not without knowledge of the domain growth law by making twin-$T$-shifts,\cite{jonyosnor2002} 
determining $\teff$  from both positive and negative $T$-shifts between two temperatures.  
The aging at a pair of temperatures ($T_1$,$T_2$) is cumulative if $L_{T_1} (t_1) = L_{T_2}(t_2)$, where $t_1=\tw$ if $T_1=T_i$ (respectively $\teff$ if $T_1=T_m$) and $t_2=\teff$ ($\tw$), while the aging is noncumulative (rejuvenated) if $L_{\Tm} (\teff) < L_{\Ti}(\tw)$.
If the aging is cumulative, in a plot of $t_1$ vs $t_2$ the two sets of data (corresponding to positive and negative $T$-shifts) fall on the same curve $t_2=f(t_1,T_1;T_2)$, where $f(t,T;T_2)=L_{T_2}^{-1}[L_T(t)]$.
We see in Fig.~\ref{teff} that data for positive $T$-shifts and negative
$T$-shifts fall onto the same curve within the errors of the estimation
of $\teff$ for $\Delta T  \leq 4$~K.  
Larger temperature shifts cannot be investigated within the experimental time window  due to the large separation in time with temperature.

Since we now know that the aging is cumulative in the investigated temperature range, we can use the ($t_1$,$t_2$) data to analyze which functional form of the domain growth law $L_T(t)$ is consistent with the data.
It can be seen in Fig.~\ref{teff} that the experimental data is well described by 
\be
t_2=\tau_0(t_1/\tau_0)^{(T_1/T_2)}\,,
\label{t2}
\ee
with  $\tau_0 \approx 4 \cdot 10^{-10}$~s.
This time-temperature relation is consistent with any thermally
activated process $t/\tau_m=\exp(B_L/k_{\rm B} T)$, where $B_L$ is the free-energy barrier originating form the dipolar interaction between the superspins. 
By setting the
free-energy barrier $B_L=k_{\rm B} \Tg{\rm ln}(L/L_0)/b$ an
algebraic growth law, 
$
L_T(t)\sim L_0(t/\tau_m)^{b T/\Tg},
$
is obtained,
as has commonly been observed in numerical simulations.\cite{komyostak99,kisetal96,maretal98}
However, $B_L=k_{\rm B} \Tg(L/L_0)^\psi$ yields a logarithmic growth
law\cite{fishus88} of the form 
$
L_T(t)\sim L_0 [(T/\Tg) {\rm ln}(t/\tau_m)]^{1/\psi}
$.
Here $\tau_m$ is the microscopic flip time of an individual magnetic
moment (superspin) and is approximately given by an Arrhenius law  
\be 
\tau_m = \tau_0 \exp(KV/k_{\rm B} T)\,,
\label{Arr}
\ee
where $KV$ is the uniaxial anisotropy barrier energy for a particle of volume $V$. 
It is assumed that the anisotropy constant $K$ only weakly depends on temperature, so that $K(T_1)\approx K(T_2)$ and hence the value of the energy barrier ($KV$) cancels out.
It should be noted that $\tau_0$ in Eq.~(\ref{t2}) corresponds to the prefactor in the Arrhenius law which is almost constant with temperature. 
We also note that Eq.~(\ref{t2}) implies that the temperature dependence
of $\Upsilon(T)$ in Eq.~(\ref{eq-prob}) is of minor importance at
the investigated temperatures (not too close to $\Tg$). 

\begin{figure*}[t!]
\includegraphics[width=0.95\textwidth]{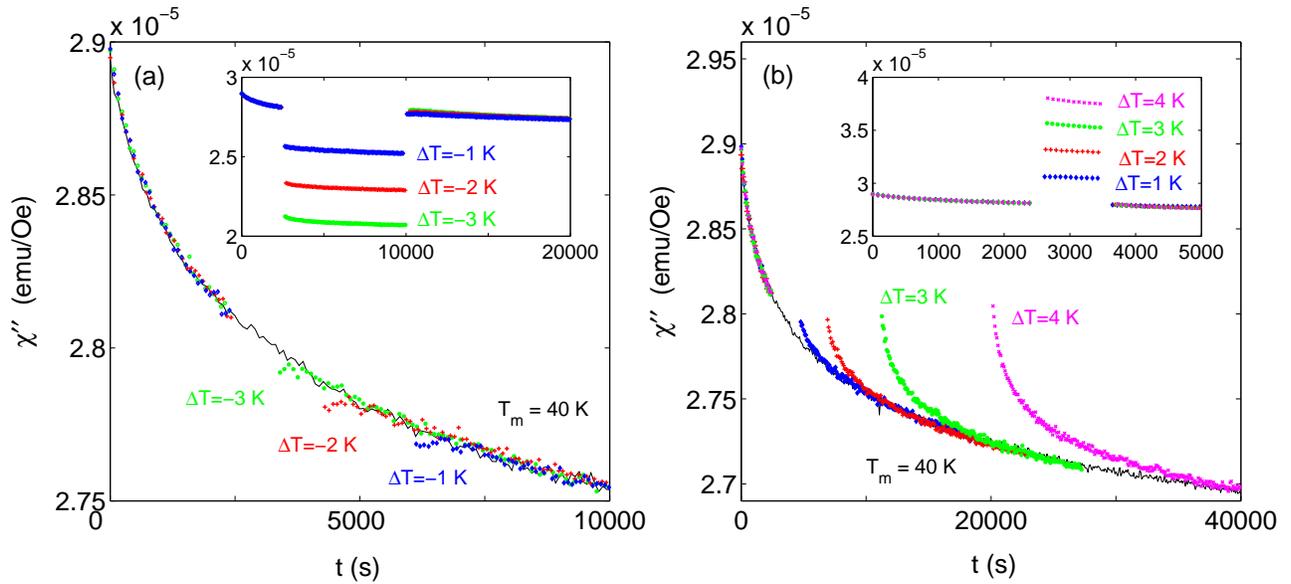}
\caption{(Color online) ac susceptibility vs time after (a) negative and (b) positive $T$-cyclings. The raw data is shown in the inset. In the main frame the relaxation data at $T_m$ after the $T$-cycling are shown shifted a quantity $\teff$ in time. $\omega/2\pi=510$~mHz. 
}
\label{ac-cycl}
\end{figure*}

For this superspin glass, the separation of time scales with temperature [Eq.~(\ref{t2})] is almost the same as for an atomic ($\tau_0=\tau_m \sim 10^{-12}$~s) Ising spin glass at temperature well below $\Tg$.\cite{dupetal2001}
This does however not mean that the time and length scales are of the same order since, for the superspin glass, the microscopic unit time is given by $\tau_m$ [Eq.~(\ref{Arr})] which is much longer than than the prefactor $\tau_0$.
It is not possible to directly probe length scales in nonequilibrium
magnetization measurements, but, a strong indication that the time (and
length) scales are shorter than in atomic spin glasses are given from
critical slowing down analysis of the critical dynamics above the
spin-glass transition.
The longest relaxation time due to correlated dynamics is given by
$
\tau_c/\tau_m \sim \xi^z\sim(T/\Tg-1)^{-z\nu}
$
where $\xi$ is the critical correlation length and $z$ and $\nu$ are critical exponents.
For superspin glasses $\tau_m$ is generally found to be much longer than the atomic spin flip time, while the value of $z\nu$ is found to be similar to those of Ising spin glasses.\cite{jon2004}
In the present system $\tau_m \sim 10^{-5} - 10^{-4}$~s was found for temperatures $T/\Tg \sim 1.3-1.7$.\cite{mamnak99}

Since the aging is cumulative in the case of $T$-shift experiments one would also expect it to be cumulative in the case of $T$-cycling experiments.
That is indeed the case, as shown in Fig.~\ref{ac-cycl}, in which
negative and positive temperature cyclings $\Tm (\twone) \to \Tm+\Delta T (\twtwo) \to \Tm$ with values of $|\Delta T|\leq 4$~K are shown.
The raw data are shown in inset and we can see that the ac susceptibility relaxes in all 3 stages of the $T$-cycling process.
In the main frame the ac data at $\Tm$ is plotted; the timescale of
$\chi''$ in the last stage (back to $\Tm$ after the $T$-cycling) is changed by an
amount of $\teff$ in order to make $\chi''$ fall on top of the reference isothermal aging curve at long time scales.
The values of $\teff$ are consistent with activated dynamics according to Eq.~(\ref{t2}) (with $\tau_0=4\cdot 10^{-10}$~s, as determined from Fig.~\ref{teff}).

\subsection{Transient relaxation}

\begin{figure}[b]
\includegraphics[width=0.45\textwidth]{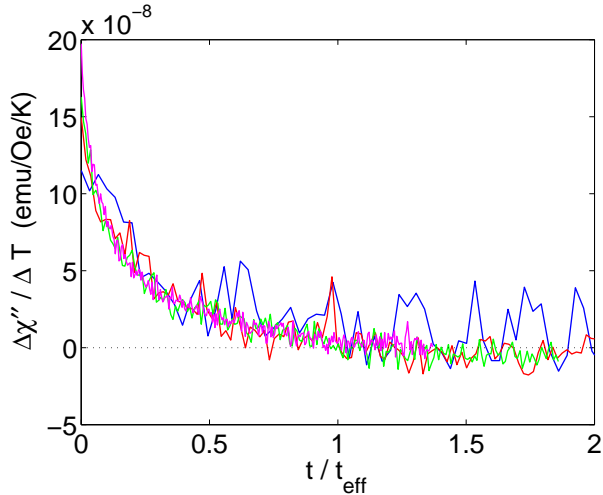}
\caption{(Color online) Scaling of the transient part of the ac susceptibility after the positive $T$-cycling shown in Fig.~\ref{ac-cycl}(b). 
$\teff$ is the effective time at $\Tm=40$~K, corresponding to the aging at $_Tm+\Delta T$. 
$\Delta \chi''$ vanishes at $t/\teff \sim 1$ confirming that the timescale for transient relaxation $\tmerge \approx \teff$.
}
\label{ac-cycl-trans}
\end{figure}

In Fig.~\ref{ac-cycl} it can be seen that at short time scales a part of the shifted $\chi''$ curve does not fit to the isothermal aging curve. We call this part {\em transient relaxation} since we know that the aging is cumulative. 
The same transient relaxation is observed both after $T$-cycling and shift experiments. We note that a large transient relaxation makes it difficult to accurately determine $\teff$. 
According to Sec.~\ref{sec-back} the transient part of the relaxation
 $\Delta \chi'' \propto \Delta T$.
In addition $\Delta \chi''$ should vanish after the time $\tmerge$ which is equal to the effective time $\teff$.
We have extracted the transient part of the ac-susceptibility $\Delta\chi''$,
for the positive $T$-cycling
experiments shown in Fig.~\ref{ac-cycl}(b),
and plotted $\Delta\chi''/\Delta T$ vs
$t/\teff$ in Fig.~\ref{ac-cycl-trans}.
All measurements with different $\Delta T$ fall on the same master curve which reaches zero at $t/\teff \sim  1$.
Hence the predictions about the transient relaxation from Sec.~\ref{sec-back} are confirmed; the amplitude of $\Delta \chi''$ is proportional to $\Delta T$ and the timescale of the transient relaxation $\tmerge \simeq \teff$.
The measurement with the smallest $\Delta T$ is the most noisy in the scaling plot since it has the smallest transient part.
We also verified that this scaling of the transient relaxation is valid if $\twone$ is varied for a fixed $\Delta T$, and that the same transient relaxation is observed after $T$-shift experiments.
After the negative $T$-cyclings shown in Fig.~\ref{ac-cycl} (as well as 
after positive $T$-shifts not shown) the amplitudes of the transient relaxation  $\Delta \chi''$ are too small  to be meaningful to make the same 
scaling analysis. 
In Fig.~4 of Ref.~\onlinecite{mamnakfur99}  transient effects are studied in negative $T$-shift experiments with $\Delta T=2$~K varying $\tw$. $\chi''$ is plotted on a logarithmic time scale and experiments with very long aging times are performed. It can clearly be seen that $\Delta \chi''$ of the transient relaxation and also $\tmerge$ becomes larger with increasing $\tw$. Also those data are consistent with the scaling proposed in Sec.~\ref{sec-back}.

\subsection{Cooling-rate effects}

\begin{figure}[htb]
\includegraphics[width=0.9\columnwidth]{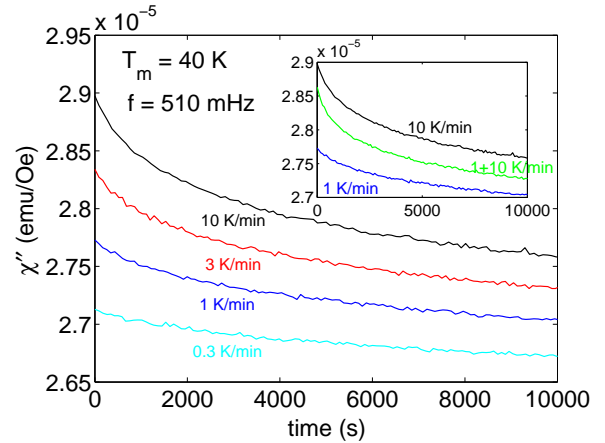}
\caption{(Color online) ac susceptibility vs time measured after cooling to 40~K with  different cooling rates. The inset show the ac relaxation after cooling with 10~K/min (upper curve) and 1~K/min (lower curve). The middle  curve is measured after cooling to 50~K with a cooling rate of 1~K/min and from 50 to 40~K with 10~K/min. $\omega/2\pi=510$~mHz. 
}
\label{coolrate}
\end{figure}

In the cumulative aging scenario, strong cooling-rate effects are expected;
the entire thermal history is important (within the spin-glass phase).
It can be seen in Fig.~\ref{coolrate} that the ac relaxation indeed depends on the cooling rate.
The system exhibits less relaxation and $\chi''$ reaches lower values after a slow cooling, which is consistent with the cumulative aging scenario---the relaxation starts from a larger effective domain size after the slower cooling than after the faster cooling process.
From these experiments it appears as if the equilibrium value of $\chi''(t\to \infty)$ changes with the cooling rate. However, if the aging is cumulative the slower relaxation and lower value of the ac susceptibility could also be explained with an effective time being well out-side the experimental time window.
The ac relaxation measured after slow cooling (1~K/min) down to 50~K and fast cooling (10~K/min) to $\Tm=40$~K is shown in the inset. This curve is different both from the measurements with slow and fast  cooling. 
It hence appears that the thermal history in the entire spin-glass phase is of importance.
These results are very different from those reported on atomic spin glasses, in which the aging at temperatures close to $\Tm$ dominates the nonequilibrium magnetic response.\cite{jonetal98,jonetal99}
For orientational glasses, on the other hand, the aging around the freezing temperature will contribute most strongly to the susceptibility at $\Tm$.\cite{albdoulev97} 
The strong cooling rate effects, in the case of orientational glasses, have been taken as evidence for ``strong ergodicity breaking''.\cite{doulevzio96,albdoulev97}

\subsection{Large temperature changes}

\begin{figure}[htb]
\includegraphics[width=0.9\columnwidth]{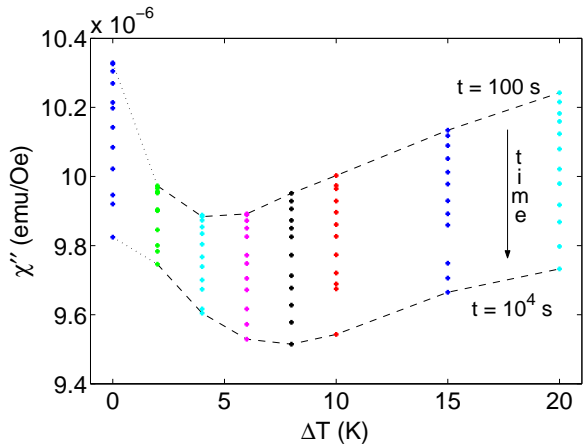}
\caption{(Color online) $\chi''$ for times 100 to $10^4$~s (up to down) at $\Tm=30$~K. 
During the cooling process a temporary stop is made at $\Tm+\Delta T$ for 1000~s.
The measurement with $\Delta T=0$ corresponds to a direct quench to $\Tm$.
The lines between $\chi''(t=100~{\rm s})$ and $\chi''(t=10^4~{\rm s})$ for different $\Delta T$'s are only a guide for the eye.
}
\label{Tshiftlarge}
\end{figure}

Since the twin-$T$-shifts experiment is limited in small $\Delta T$ due to
the large separation in time with temperature, we look for possible
rejuvenation effects by other temperature-change protocols involving 
larger $\Delta T$. 
In Fig.~\ref{Tshiftlarge} large negative $T$-shift experiments are shown
for various values of $\Delta T$ in order to test at which temperature
during the cooling process that the aging is the most efficient. 
It can be seen that if the intermittent stop of the cooling is made
close to the target temperature $\chi''$ reaches lower values with increasing $\Delta T$. This
is interpreted by the cumulative aging scenario, similarly to the
cooling-rate effect discussed above. 
However, for yet larger values of $\Delta T$ ($\gtrsim 8$~K),
 the relaxation increases
and $\chi''$ again becomes larger. This nonmonotonic behavior of
$\chi''(t)$ indicates that the aging effect might  not be completely
cumulative in the entire spin-glass phase. 
On the other hand, assuming that the aging is cumulative, for the measurement with $\Delta T = 20$~K, $\tmerge$ is of the order of $10^{11}$~s. 
It is hence impossible to say if the curve that ``appears to be
rejuvenated'' reveals in fact the rejuvenation due to temperature-chaos
or the ``fixed energy-landscape rejuvenation'' discussed in
Sec.~\ref{sec-back}. In the latter picture, the large droplets created
at the higher temperature are completely frozen in at the lower
temperature on the experimental time window. They would be {\em
deactivated} only at much longer timescales ($\tmerge \approx \teff$).

\begin{figure}[htb]
\includegraphics[width=0.9\columnwidth]{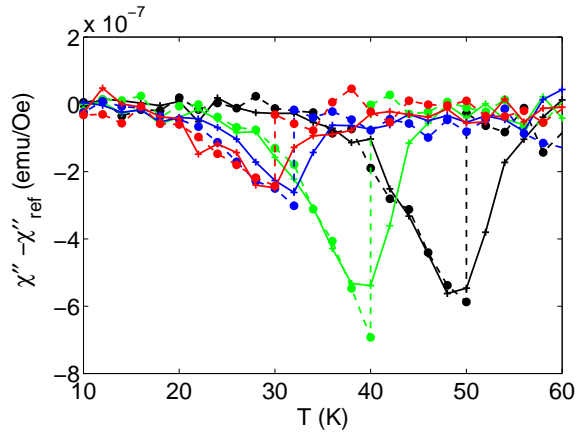}
\caption{(Color online) $\Delta\chi''$ vs temperature measured on cooling (circles connected by dashed lines) and reheating (pluses connected by solid lines).
A temporary stop is made on cooling  at $\Ts=50$, 40, 32, or 30~K for $\ts=9000$~s. $\omega/2\pi=510$~mHz.}
\label{mem}
\end{figure}

The effect of large temperature changes can also be investigated by so-called memory experiments;\cite{jonetal98} the ac susceptibility is measured on cooling with a temporary stop at a temperature $\Ts$. After aging the sample for a time $\ts$, the cooling is resumed. The ac susceptibility is subsequently measured also on reheating.
Such measurements with $t_s=9000$~s and $\Ts=50, \,40,\, 32,$ and 30~K are shown in Fig.~\ref{mem}.
After the aging at $\Ts$ the ac susceptibility slowly merges with the reference curve. In a similar way, $\chi''$ comes closer to the isothermal reference curve in the $T$-shift experiments shown in Fig.~\ref{Tshiftlarge}.
The wide (asymmetric) memory dips can be explained by freezing of large domains (grown during the aging at $\Ts$) on cooling and unfreezing of these domains on reheating, without involving the concept of $T$-chaos.

\section{Discussion}

\begin{table*}[t]
\caption{Comparison of the effects of temperature changes in the
 superspin glass and atomic spin glasses.
Cooling rate effects are always observed in the region of cumulative aging. The absence of cooling rate effects observed for the atomic Heisenberg spin glasses therefore refer to the cooling rate not too close to the target temperature.
In ac-memory experiments, the duration of the halt, as well as the cooling/heating rate employed affect the results. The width of the memory dip is therefore only a qualitative measure of the rejuvenation effect.  
\label{table}}
\vspace{2mm}
\begin{tabular}{|c|c|c|c|c|}\hline
 & superspin glass & atomic Ising SG & 
\multicolumn{2}{c|}{atomic Heisenberg SG} \\
\cline{4-5}
  & Fe$_3$N superspin glass 
& Fe$_{0.5}$Mn$_{0.5}$TiO$_3$~\cite{ghoststory,jonyosnor2002,dupetal2001} & 
Ag(11 at\% Mn)\cite{ghoststory,jonyosnor2002} & 
CdCr$_{1.7}$In$_{0.3}$S$_4$~\cite{dupetal2001,sasetal2002} \\ \hline
$\Tg$  & $\sim$ 60~K & 21.3~K & 32.8~K & 16.7~K \\ \hline 
twin-$T$-shift around $T_m$ & 36--40~K & 17.5--19~K & 29.5--30~K & not  investigated  \\
cumulative aging & $|\Delta T| \nle 4.0$~K & $|\Delta T| \nle 0.5$~K & 
$|\Delta T| \nle 0.05$~K & \\
noncumulative aging & & & & \\
({\it rejuvenation}) & not observed & $|\Delta T| \nge 1.5$~K & 
$|\Delta T| \nge 0.2$~K &  \\ \hline
cooling-rate & & & & \\
dependence & yes & not investigated & no & no \\ \hline
width of & & & & \\
dip in $\Delta \chi''$ & $\sim$ 10~K & $\sim$ 3~K & $\sim$ 2~K & 
$\sim$ 2~K \\ \hline
\end{tabular}
\end{table*}

The glassy dynamics of this superspin glass, with random uniaxial
anisotropy and dipolar interaction, is very similar to that observed in
numerical simulations on the 3d Ising EA
model;\cite{takhuk2002,komyostak2000A} the 
aging is cumulative up to relatively large values of $\Delta
T/\Tg$. Even for the largest possible temperature differences in the
twin-$T$-shift analysis rejuvenation due to $T$-chaos cannot be observed. 
In a $T$-shift process with cumulative aging, the domains grown at the first temperature $T+\Delta T$ are still relevant after the $T$-shift.
At the new temperature $T$, the system appears to have been aged for an effective time $\teff$, fulfilling  $L_{T}(\teff)=L_{T+\Delta T}(\tw)$.
However,  a characteristic transient relaxation is observed after the $T$-shift
and   disappears after a finite time $\tmerge \approx \teff$.
The aging dynamics in the transient
range ($t<\tmerge$) can be understood as the turning on/off of thermally active
droplets.  It can also be interpreted {\it mutatis mutandis} as the
Kovacs effect (in a two-time variable) that has earlier been observed in
structural glasses\cite{kovacs63,kovetal79} and numerical simulations on
spin glasses.\cite{berbou2002,komyostak2000A} 
As increasing $\Delta T$
in a negative $T$-shift, the temperature $\teff$ becomes very long due
to the separation of time scales with temperature.  
If $\teff$ falls outside the observation time window, the transient relaxation can be interpreted as ``rejuvenation in  fixed-energy landscape''.\cite{bouetal2001,berbou2002}
This effect is intrinsically different from {\em rejuvenation due to
temperature chaos} in the respect that after a finite time ($\tmerge$)
the cumulative nature of the aging is revealed.  
However within a limited time window it is difficult to distinguish the
transient effect from the rejuvenation (chaos) effect.

Let us now compare the effects of temperature changes observed by the
experiments studied  on the present sample with those of atomic
spin glasses (c.f. Table.~\ref{table}).  In the atomic Heisenberg spin
glass Ag(11 at\% Mn) the aging judged from twin-$T$-shift experiments is
cumulative only for quite small $\Delta T$. Interestingly, the range
of $\Delta T$ where the cumulative aging is ascertained in the atomic Ising
spin glass Fe$_{0.5}$Mn$_{0.5}$TiO$_3$ is larger than that of Ag(11 at\% Mn) by an
order of magnitude, but it is still smaller than that of  the presently studied  superspin
glass  by another order of magnitude. Correspondingly, 
$\Delta T$'s for which noncumulative memory (rejuvenation) is clearly
recognized behave similarly, though for the superspin glass no evidence for noncumulative aging were found (within our experimental time window).
In a memory experiment with an intermittent stop on cooling,
the width of the dip in $\Delta \chi''$ gives only a quantitative measure of 
the rejuvenation effect, but it can be noticed that it is  relatively larger for the superspin glass than for the atomic spin glasses. Large memory dips have also been observed in earlier investigations on different superspin glass systems.\cite{jonhannor2000,sahetal2003}
 We can thus
read from Table~\ref{table} that Heisenberg spin glasses are very sensitive to
temperature changes as already pointed 
out,\cite{dupetal2001,jonyosnor2002,ghoststory} and that the effects of
temperature changes in the superspin glass  are qualitatively rather
similar, although weaker,  to those of the atomic Ising spin glass Fe$_{0.5}$Mn$_{0.5}$TiO$_3$. 

In the droplet picture, rejuvenation effects will appear on length
scales larger than the overlap length $L_{\Delta T}$.  
While the noncumulative aging and the strong rejuvenation effects
observed in atomic spin glasses can be
attributed to the influence of temperature-chaos on the investigated
length scales, the cumulative aging, the cooling-rate dependence, and
even the dip in $\Delta \chi''$ in the memory experiment of the present
system can be explained by the much shorter length scales probed being
much smaller than $L_{\Delta T}$. 
The reason for the different length scales probed in Ising atomic and
superspin glasses within the same experimental time window is attributed to the
difference in microscopic timescale ($\tau_m \sim 10^{-12}$~s for atomic
spins while $\tau_m > 10^{-5}$~s for the here investigated superspins).
In addition the nanoparticle sample is Ising-like due to uniaxial anisotropy and rejuvenation effects related to temperature-chaos have been shown to be weaker in Ising systems than in Heisenberg systems.\cite{ghoststory,krz2003}

\section{Conclusion}

We have studied the effect of temperature changes on the nonequilibrium dynamics of a superspin glass (a strongly interacting nanoparticle system).
The sample exhibits cooling-rate effects and the aging after a $T$-shift can be described as the sum of a cumulative part and a transient part.
The transient relaxation can within  a real space model (the droplet model) be understood as the change of thermally active droplets associated with the $T$-shift. This transient relaxation resembles the Kovacs effect observed in various glassy systems.
Strong rejuvenation effects, as those observed in atomic spin glasses, have not been observed in this superspin glass.
According to the droplet model, this indicates that all length scales investigated in the experimental time scale are shorter than the so-called overlap length. 
This is attributed to the fact that, the microscopic flip-time of the magnetic moments (superspins) are much longer than that of an atomic spin and the length scales investigated within the experimental time window are thus much shorter.

\acknowledgments
We thank Munetaka Sasaki for stimulating discussions. 
 P.E.J. acknowledge financial support from the Japan Society for the Promotion of Science.
The present work is supported by a Grant-in-Aid for Scientific Research
Program (\# 14540351) and NAREGI Nanoscience Project from the
Ministry  of Education, Science, Sports,
Culture and Technology.

\end{document}